\def \SAIT #1 #2 {{\em Mem.\ Soc.\ Astron.\ It.\/} {\bf #1}, #2}
\def \MESS #1 #2 {{\em The Messenger\/} {\bf #1}, #2}
\def \ASTRNACH #1 #2 {{\em Astron. Nach.\/} {\bf #1}, #2}
\def \AAP #1 #2 {{\em Astron. Astrophys.\/} {\bf #1}, #2}
\def \AAL #1 #2 {{\em Astron. Astrophys. Lett.\/} {\bf #1}, L#2}
\def \AAR #1 #2 {{\em Astron. Astrophys. Rev.\/} {\bf #1}, #2}
\def \AAS #1 #2 {{\em Astron. Astrophys. Suppl. Ser.\/} {\bf #1}, #2}
\def \AJ #1 #2 {{\em Astron. J.\/} {\bf #1}, #2}
\def \ANNREV #1 #2 {{\em Ann. Rev. Astron. Astrophys.\/} {\bf #1}, #2}
\def \APJ #1 #2 {{\em Astrophys. J.\/} {\bf #1}, #2}
\def \APJL #1 #2 {{\em Astrophys. J. Lett.\/} {\bf #1}, L#2}
\def \APJS #1 #2 {{\em Astrophys. J. Suppl.\/} {\bf #1}, #2}
\def \APSS #1 #2 {{\em Astrophys. Space Sci.\/} {\bf #1}, #2}
\def \ASR #1 #2 {{\em Adv. Space Res.\/} {\bf #1}, #2}
\def \BAIC #1 #2 {{\em Bull. Astron. Inst. Czechosl.\/} {\bf #1}, #2}
\def \JSQRT #1 #2 {{\em J. Quant. Spectrosc. Radiat. Transfer\/} {\bf #1}, #2}
\def \MN #1 #2 {{\em Mon. Not. R. Astr. Soc.\/} {\bf #1}, #2}
\def \MEM #1 #2 {{\em Mem. R. Astr. Soc.\/} {\bf #1}, #2}
\def \PLR #1 #2 {{\em Phys. Lett. Rev.\/} {\bf #1}, #2}
\def \PASJ #1 #2 {{\em Publ. Astron. Soc. Japan\/} {\bf #1}, #2}
\def \PASP #1 #2 {{\em Publ. Astr. Soc. Pacific\/} {\bf #1}, #2}
\def \NAT #1 #2 {{\em Nature\/} {\bf #1}, #2}

\documentstyle{memsait}
\input epsf.sty
\begin{opening}
\title{SLOTT-AGAPE Project\footnote{Italian Component.}}
\author{V. Bozza$^3$, S. Calchi Novati$^3$,  M. Capaccioli$^5$, S. Capozziello$^3$, \\
V. Cardone$^2$, G. Covone$^2$, F. De Paolis$^1$, R. de Ritis$^2$,
G. Ingrosso$^1$,\\ G. Iovane$^3$, G. Lambiase$^3$, A.A.
Marino$^5$, G. Marmo$^2$, I. Musella$^5$, \\ E. Piedipalumbo$^2$,
S. Pezzuto$^1$, M. Roncadelli$^4$, C. Rubano$^2$, \\ G.
Scarpetta$^3$, P. Scudellaro$^2$, F. Strafella$^1$}
\institute{$^1$Universita' di Lecce, Italy\\ $^2$Univesita'
Federico II, Napoli, Italy\\$^3$Universita' di Salerno,
Italy\\$^4$INFN, Pavia, Italy\\$^5$Osservatorio Astronomico di
Capodimonte,Napoli, Italy}
\date{} 
\end{opening}

\begin{document}

\oddpagefooter{}{}{} 
\evenpagefooter{}{}{} 
\
\bigskip

\begin{abstract}
SLOTT-AGAPE (Systematic Lensing Observation at Toppo Telescope -
Andromeda Gravitational Amplification Pixel Lensing Experiment) is
a new collaboration project among international partners from
England, France, Germany, Italy and Switzerland that intends to
perform microlensing observation by using M31 as target. The
MACHOs search is made thanks to the pixel lensing technique.
\end{abstract}

\section{Introduction}
In recent years, much attention was centred on the possibility
that dark matter, and in particular its baryonic side, consists of
astrophysical objects, generically termed as "Massive
Astrophysical Compact Halo Objects" (MACHOs) with mass
$10^{-8}M_{\odot }<M<10^{-2}M_{\odot }$ . Direct searches of these
objects can, at best, reach the solar neighborhood. In order to
detect them further out, it was proposed by Paczynski (1986) to
search for dark objects by gravitational microlensing [1].

Microlensing is an application of General Relativity effect of
gravitational lensing where the separation between the produced
images is too small to be appreciated ($\delta\theta\le 10^{-3}$
arcsec); nevertheless -- owing to the motion of the lens -- it
produces a time--dependent light amplification of the source which
is observable. In fact, when a compact object passes nearby the
line of sight of a background star, the luminosity of this star
increases giving rise to a characteristic luminosity curve (see
Fig.1).

The galactic structure is not very well understood due to the
ignorance of the effective content of dark matter and its actual
distribution, so the first goal is to perform a map of the MACHOs'
dark matter distribution   both in the galactic disk through
microlensing observations  towards the galactic bulge\footnote{Of
course, observations  towards the galactic bulge are not easy from
the Northern Hemisphere, but possible.} and the spiral arms
(Sagittarius, $\gamma$Scuti, $\beta$Scuti) and in the galactic
haloes of the Milky Way, M31, M33 and dwarf galaxies.

Until now a lot of microlensing events have been detected towards
the galactic bulge and the LMC [2][3]. These results have allowed
to better understand the galactic structure. For example it has
been recently suggested that the bulge has not a simple spherical
symmetry but it has also a barred structure [4]. Microlensing
searches by MACHO and EROS groups, looking for the magnification
of LMC stars by MACHOs, have now been underway for several years.
The very low microlensing probability requires several millions
stars to be daily monitored, in order to observe significant
luminosity increases. Some events have been reported, but less
than expected in the standard halo model. Other experiments (DUO
and OGLE), as well as MACHO itself, are monitoring stars of the
Galactic bulge in order to look for microlensing by stars in the
Galactic disk and in the bulge itself. These appear to find more
events than expected [5].

The AGAPE and Columbia-Vatican (VATT) collaborations look for
microlensing in the direction of the M31 galaxy. This could yield
very useful information on the haloes of both our own Galaxy and
M31. A pilot observation at Pic-du-Midi Observatory by AGAPE has
given promising results [6].

Our project is to extend observations towards all the visible
targets from the Potenza Toppo Observatory in order to detect a
larger amount of microlensing events.
\begin{figure}[h!]
\epsfysize=5,5cm 
\hspace{1.5cm}\epsfbox{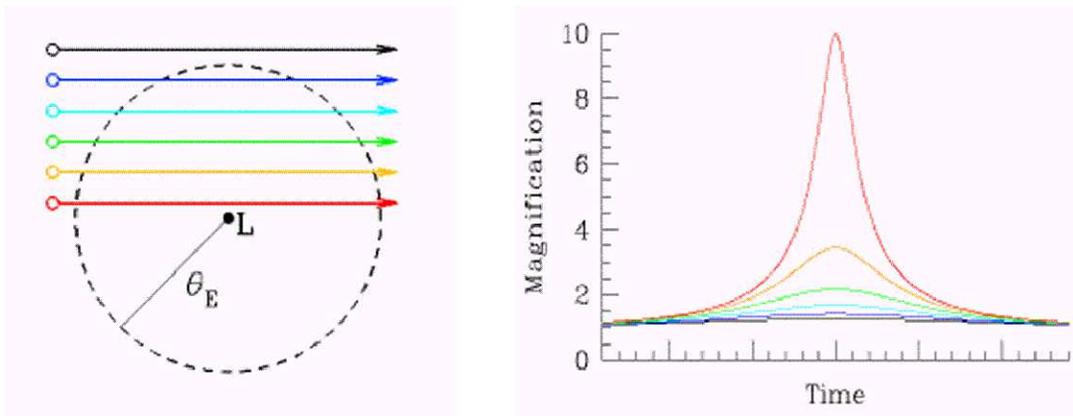} 
\caption[h]{Characteristic luminosity curve and relared impact
parameter.}
\end{figure}

Another goal is to search for planets or binary lensing events for
which a very accurate photometry is required. Toppo Telescope has
the right features to contribute substantially to this aim. This
fact will also allow us to investigate in detail the initial
stellar mass function as well as the presence of planets (both
Jupiter--like and Earth--like) [7].

\section{Gravitational lensing}
The light deflection due to a gravitational field ( in weak field
approssimation and geometrical optics approssimation) is described
by lens equation
\begin{equation}
\mathbf{\eta }=\frac{D_{s}}{D_{d}}\mathbf{\xi }-D_{ds}\widehat{\mathbf{%
\alpha }}(\mathbf{\xi }),
\end{equation}

with
\begin{equation}
\widehat{\mathbf{\alpha }}(\mathbf{\xi })=\int d^{2}\mathbf{\xi }^{\prime }%
\frac{4G\Sigma (\mathbf{\xi }^{\mathbf{\prime }})}{c^{2}}\frac{\mathbf{\xi }-%
\mathbf{\xi }^{\mathbf{\prime }}}{\left| \mathbf{\xi }-\mathbf{\xi }^{%
\mathbf{\prime }}\right| }
\end{equation}

and $D$ is euclidean distance on galactic scale, while angular
diameter distance on extragalactic scale.

The Schwarschild one is the simplest lens system, made of a
point-like deflector. In this case, the deflection angle is:

\begin{equation}
\widehat{\alpha }=\frac{4GM}{c^{2}b},
\end{equation}

where $M$ is the deflector mass and $b$ the impact parameter.
\section{Pixel Lensing}
In a dense field, many stars contribute to any pixel of the CCD
camera at the focal point of the telescope. If an unresolved star
is sufficiently magnified, the increase of the light flux can be
measured on the pixel. Therefore, instead of monitoring individual
stars, we follow the luminous intensity of the pixels. Then all
stars in the field, and not the only few resolved ones, are
candidates for a microlensing event; so the event rate is
potentially much larger. Of course, only the brightest stars will
be amplified enough to become detectable above the fluctuations of
the background, unless the amplification is very high and this
occurs very seldom. In a galaxy like M31, however, this is
compensated by the very high density of stars.

The first step for the analysis of light curves is to define the
baseline or the background flux $\phi _{background}$. This is done
by taking the minimum of a sliding average on 10 consecutive
points. One can define the beginning of a "bump"\footnote{A
significative variation of luminosity on an opportune group of
pixels connected with the dimension of the average PSF.}  if 3
consecutive points lie 3$\sigma $ above $\phi _{background}$ and
the end when 2 consecutive points fall below 3$\sigma $. The
second step is to select microlensing candidates by light curves
with only one bump and not more. The third step is to fit a high
amplification degenerate Paczynski curve to the mono-bumps. The
amplification is then well approximated by

\begin{equation}
A(t)-1\approx \frac{1}{u(t)}\qquad with\qquad u(t)=\sqrt{\left( \frac{t-t_{0}%
}{t_{e}}\right) ^{2}+u_{0}^{2}},
\end{equation}

where the Einstein time is  $t_{e}=R_{e}/V_{\bot }$  , the ratio
of the Einstein radius to the transverse velocity of the lens.

In Fig.2 we find a typical light curve obtained with AGAPE method.
\begin{figure}[h!]
\epsfysize=5,5cm 
\hspace{1.5cm}\epsfbox{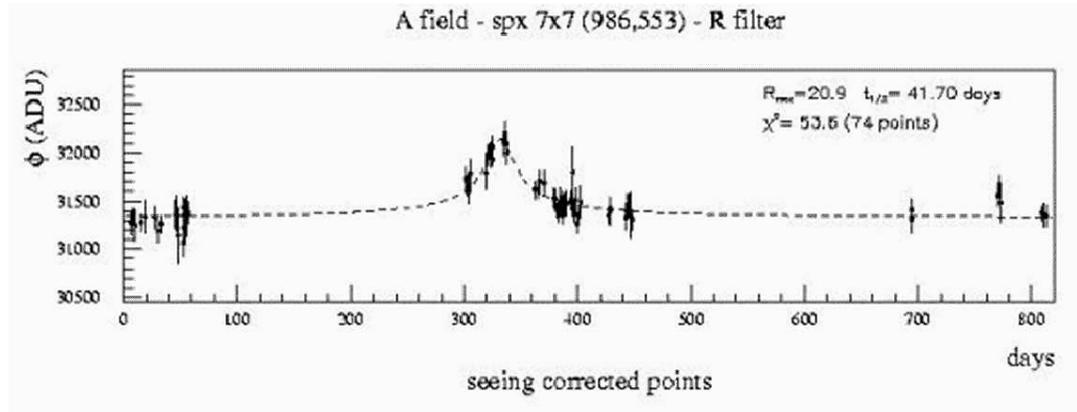} 
\caption[h]{Characteristic Paczynski luminosity curve.}
\end{figure}

\section{Conclusion}
The achievement could allow us to obtain:\\ a) the first large
microlensing survey performed in the Northern Hemisphere for
spiral arms observations and marginally for the bulge by taking
into account the geographic position, the new generation optics
and device at the Toppo  telescope;\\ b) a detailed survey on
other galaxies besides the Galaxy (first of all M31);\\ c) the
capability of detecting planets (both Jupiter-like and
Earth-like);\\ d) the partecipation in the follow-up observations
of microlensing events which will be announced by program like the
Global Microlensing Alert Network (GMAN) or Planet
Collaboration;\\ e) the possibility to use larger or spacecraft
telescope as HST to resolve interesting pixels obtaining more
astrophysical information on the amplified objects.

An on-line selection and a quasi-on-line analysis could be made to
performe the last point [8].

\end{document}